\documentclass[aps,prl,twocolumn,floatfix]{revtex4-2}
\usepackage{boldtensors}
\usepackage{graphicx}

\begin{document}

% Use the \preprint command to place your local institutional report
% number in the upper righthand corner of the title page in preprint mode.
% Multiple \preprint commands are allowed.
% Use the 'preprintnumbers' class option to override journal defaults
% to display numbers if necessaryThe relaxation dynamics of an epidemic
%\preprint{}

%Title of paper
%\title{A new tool to determine the large-scale evolution of an epidemic}
%\title{Large-scale entropic decay of an epidemic towards Herd immunity}
\title{Entropic decay of an epidemic towards herd immunity}
%\title{Large-scale primary decay of an epidemic by entropic effects towards Herd immunity}
% repeat the \author .. \affiliation  etc. as needed
% \email, \thanks, \homepage, \altaffiliation all apply to the current
% author. Explanatory text should go in the []'s, actual e-mail
% address or url should go in the {}'s for \email and \homepage.
% Please use the appropriate macro foreach each type of information

% \affiliation command applies to all authors since the last
% \affiliation command. The \affiliation command should follow the
% other information
% \affiliation can be followed by \email, \homepage, \thanks as well.
\author{L. Vanel}
\email[]{loic.vanel@univ-lyon1.fr}
\affiliation{Institut Lumi\`ere Mati\`ere, Univ Claude Bernard Lyon 1, Univ Lyon, CNRS; F-69622, Villeurbanne, France.}

% single time scale compared to the simplest epidemic models (SIR)
% difference between infected and infectious
% data show infected
% elements that show the importance of mixing entropy : the correct epidemic shape is obtained when ce=0.5 and the exponential relaxation of ln N
% even in inhomogeneous systems one can expect exponential relaxation of entropy
% many different competing models, but none takes into account exponential relaxation of ln N?
% empirical rate laws are still used in advanced thermodynamic theories of chemical reactions
% Van't hoff rate laws
% Onsager-like rate laws predict a linear relation between reaction rate and affinity, but without the additional dependences on concentrations found in experimantal rate laws or Van't hoff rate laws.
% Collision theories explain the proportionality of rate laws with concentration of reactants
%

\date{\today}

% \begin{abstract}
% Describing the growth of an epidemic at the scale of individuals has led to increasingly complex epidemic models. Nevertheless, epidemic waves often have a similar shape at large scales that even the simplest models seem to capture. We show here that, since the very start of an epidemic wave, there is an exponential decay mechanism occurring simultaneously with epidemic growth that has not been properly accounted for. Using out-of-equilibrium thermodynamics, we build a model in which entropic effects are responsible for the observed exponential decay towards a stable infected population fraction. We find that the elementary probability of being infected determines this fraction, leading to a thermodynamic criterion for herd immunity and epidemic outbreaks.
% \end{abstract}
\begin{abstract}
The shape of an epidemic wave in simple epidemic models applies to a homogeneous distribution of infected people in the population. In large inhomogeneous systems, at country-scale for instance, the wave shape is similar except for the short time behavior. For such cases, we show that the full wave shape is tied to an exponential decay. Using out-of-equilibrium thermodynamics, we build a model in which this decay results from an increase in entropy until reaching a stable infected population fraction. We find that the elementary probability of being infected determines this fraction, leading to a thermodynamic criterion for herd immunity and epidemic outbreaks.
\end{abstract}
% insert suggested keywords - APS authors don't need to do this
%\keywords{}

\maketitle

During the COVID-19 pandemic, scientists have been pressed with questions on the duration of an epidemic outbreak, the moment when the peak will be reached and the wave starts to slow down, or when to release public health restrictions without triggering another outbreak. In practice, reducing contact with infected people, barrier measures against disseminated viruses and vaccination have been the main strategies to try and control the circulation of the SARS-CoV-2 virus. As a result, many of the COVID-19 epidemic waves have strongly decelerated before the population reaches a collective level of herd immunity, making the population prone to multiple epidemic waves \cite{Fontanet2020}. There is an inherent complexity in taking into account all the relevant parameters of an epidemic, such as virus contagiousness, incubation time, infection duration, age-dependent effects, heterogeneous population density and spatial dynamics \cite{Noble1974, Ferguson2006,Britton2020}. However, epidemic waves often have a similar shape, regardless of the size of the regions or countries considered and the public health strategies implemented in those countries.
%This suggests that, despite the complexity of the process, epidemic growth displays self-similar properties when considering large enough scales.
%If the shape of an epidemic wave was simple enough to predict, it could be used in principle to  make early projections of the evolution of an epidemic wave, including the arrival of the peak and the overall duration of the wave.

\begin{figure}[t!]
\centerline{\includegraphics[width=8.5cm]{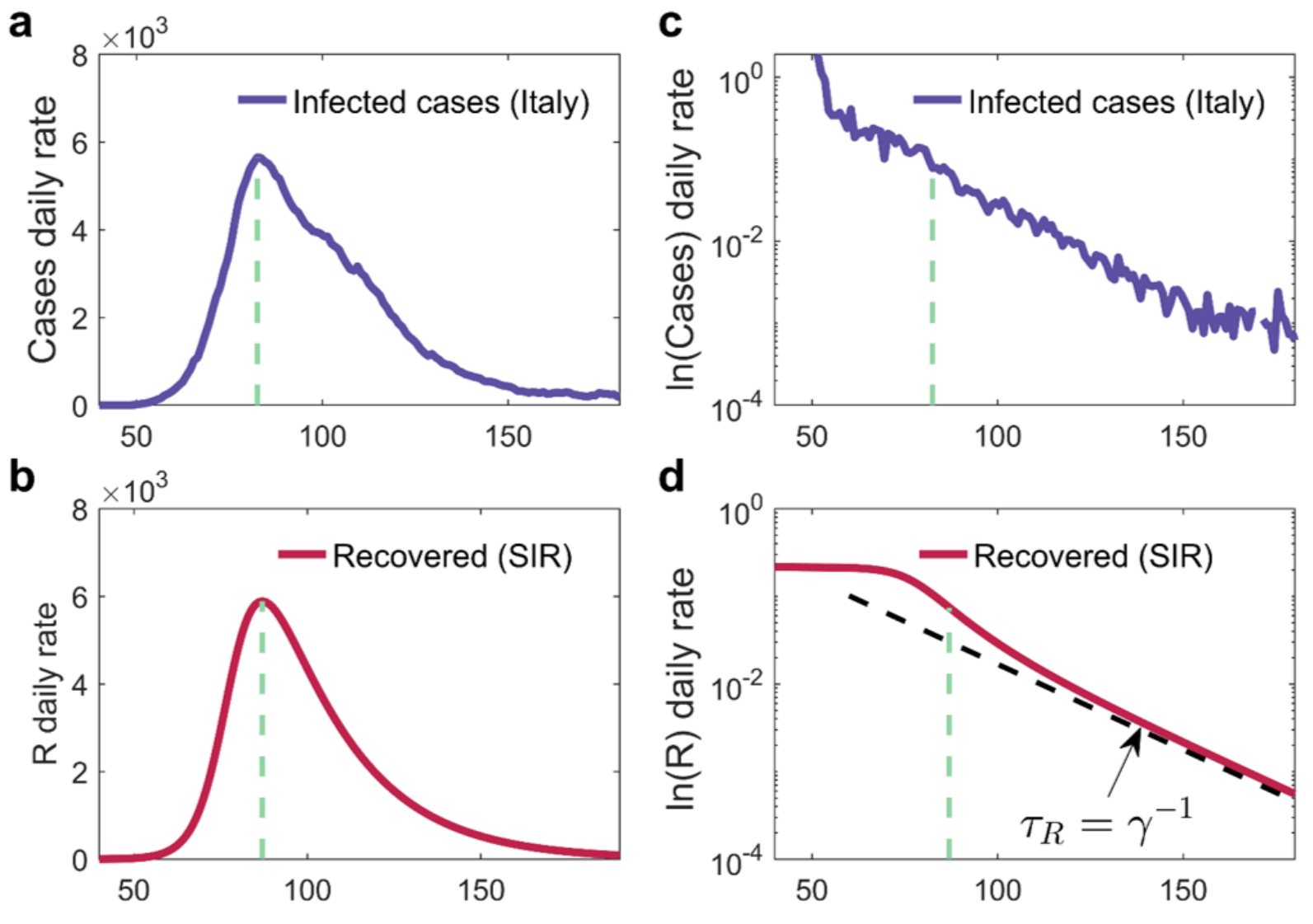}}
\caption{\textbf{Epidemic data and the SIR model.} (a), Daily rate of infections in Italy during the first COVID-19 wave. (b) The recovered population daily rate in the SIR model qualitatively agrees with epidemic data. (c), The daily rate of the logarithm of the infected population in Italy displays an almost purely exponential time relaxation. (d), The daily rate of the logarithm of the recovered population in the SIR model displays an exponential time relaxation determined by the recovery rate $\gamma$ (black dashed line) only passed the time that the peak rate has been reached in (b). Vertical dashed lines indicate the time of the peak rate in (a) and (b). }
\label{FigDataSIR}
\end{figure}

The most well-known models that predict the shape of an epidemic wave are the compartmented epidemic models \cite{Murray2002}. The simplest one is the SIR model, where S stands for susceptible people who are not yet infected, I stands for infectious people who may transmit the virus and R people who have recovered and are noninfectious \cite{Murray2002}. It is worth noting that following the evolution of population I of infectious people is not the same as counting the people who have been infected, as reported daily during the COVID-19 pandemic \cite{owidcoronavirus}. The number of infectious people I increases with the arrival of newly infected people, but decreases when people recover, or in the worst case pass away. Hence, the infectious population is a transitory state that starts from zero, reaches a peak and returns to zero. On the other hand, the infected/contaminated population $C$ continues to grow, and its daily variations produce a characteristic asymmetric epidemic shape (Fig.~\ref{FigDataSIR}a). In the SIR model, the quantity that reproduces qualitatively well the epidemic wave shape is the daily rate of recovered population (Fig.~\ref{FigDataSIR}b). Such a correspondence makes sense since there is mainly a time-shift between the daily rate of newly infected people and the daily rate of the recovered population determined by the average time for recovery. However, a lesser agreement is obtained when considering the logarithm of the total infected population $\ln C$ and its daily variations (Fig.~\ref{FigDataSIR}c) and comparing it to the daily variations of the logarithm of the recovered population (Fig.~\ref{FigDataSIR}d). Indeed, an almost pure exponential time relaxation is observed in epidemic data ($d \ln C/d t \sim \exp(-t/\tau)$), but this exponential behavior is captured by the SIR model only at times beyond the peak of the epidemic wave (vertical dashed lines), with a relaxation time $\tau_R=\gamma^{-1}$ equal to the inverse of the recovery rate $\gamma$ ($d \ln R/d t \sim \exp(-t/\tau_R)$) \cite{SupMat}. The initial plateau in the SIR model means that the rate of exponential growth at the start of the epidemic is constant for a long time, while epidemic data suggest that this rate is already decreasing exponentially over time. In the hope of better describing the early shape of an epidemic wave, phenomenological approaches have introduced ad hoc quantities that decrease exponentially with time since the start of an epidemic outbreak, either in the framework of a compartmented epidemic model \cite{Ohnishi2020,Lanteri2020} or through the use of an empirical fit based on Gompertz functions \cite{Nutter1997,Buerger2019,Utsunomiya2020}, but both approaches lack scientific justifications.
%There is no scientific basis for these empirical approaches, except for their qualitative compatibility with the steady exponential decrease in the logarithm of the infected cases daily rate (Fig.~\ref{FigDataSIR}c).

%In other words, epidemic data suggest that there is an exponential relaxation mechanism at play since the start of an outbreak that competes with the well-known initial exponential growth and causes an epidemic wave to eventually reach a peak, slow down and disappear at long times.  Exponential relaxation mechanisms are often a sign that an out-of-equilibrium system is converging towards a stable state, with fast dynamics far away progressively slowing down as the system approaches this state. However, then, in the case of an epidemic, which physical quantity could vary at an exponentially slowing-down rate and converge towards a stable state since the very start of an epidemic outbreak?

In this article, we describe a new thermodynamic analogy between an epidemic and chemical reactions that gives physical significance to the exponential relaxation observed in epidemic data since the start of an outbreak. We consider here a binary reactive mixture where one of the molecular species would represent the susceptible population S and the other one the infected population C, whether still infectious or not. During the reaction, infectious population I is considered a chemical intermediate, which is necessary for the reaction to occur but is a temporary state of the population that does not contribute to the final ``thermodynamic equilibrium'' between the susceptible and infected population. The thermodynamic equilibrium of binary mixtures is determined by the equality between the chemical potential of the two molecular species. The chemical potential has two contributions: the first one is the interaction energy associated with the addition of a molecule in a pure molecular solution; the second one takes into account the effect of molecules being dispersed in a binary mixture. For an ideal binary mixture, this term is proportional to the logarithm of the molecular fraction. Hence, when the molecular fraction of one molecular species is far from equilibrium, so will be its chemical potential. Considering the population as a mixture of people who have already been infected and people who have not yet, the binary mixture contribution to the chemical potential of the infected population C would be proportional to the logarithm of C. Initially, the binary mixture contribution of the infected population would be far from the one associated with the susceptible population. However, as the number of infected persons increases, the difference in chemical potential due to the binary mixture contribution would progressively disappear, as the fraction of infected people increases and the fraction of susceptible people decreases. This process would stop when the equality of the chemical potentials is reached at which point  there is thermodynamic equilibrium. Thus, in this analogy, the exponential decay observed for an epidemic from the start to the end of the wave would correspond to an exponential relaxation of the chemical potential towards some stable state value.

To better understand the meaning of this analogy for an epidemic, it should be remembered that mixing entropy determines the chemical potential dependence on molecular fraction. Mixing entropy is a probabilistic quantity that measures the number of possibilities to spatially distribute molecules for a given composition of a molecular mixture. In the case of an epidemic, it would correspond to the logarithmic measure of the number of ways to pick infected persons in a population of susceptible persons. In a binary mixture, mixing entropy starts from zero for the pure molecular solution, increases with the added fraction of a second molecular species and reaches a maximum when this fraction is half of the binary mixture. Generally, an increase in entropy in out-of-equilibrium systems is a driving mechanism for its evolution because it contributes to decreasing the free energy of the system. This happens because states with higher entropy have a higher probability to exist than states with low entropy. For an energetically unfavorable reaction, an increase in entropy makes it possible to happen, but would lead to an incomplete reaction with a final state that may still contain a nonnegligeable fraction of the reactants. When the chemical potential of the two species in their pure molecular state is the same, the evolution of the binary mixture will be controlled only by mixing entropy. In that case, the stable state will be half and half of each molecular species. Beyond this point, the decrease in mixing entropy means that, from a probabilistic point of view, entropy can no longer help to increase the infected population.

\begin{figure}[]
\centerline{\includegraphics[width=8.5cm]{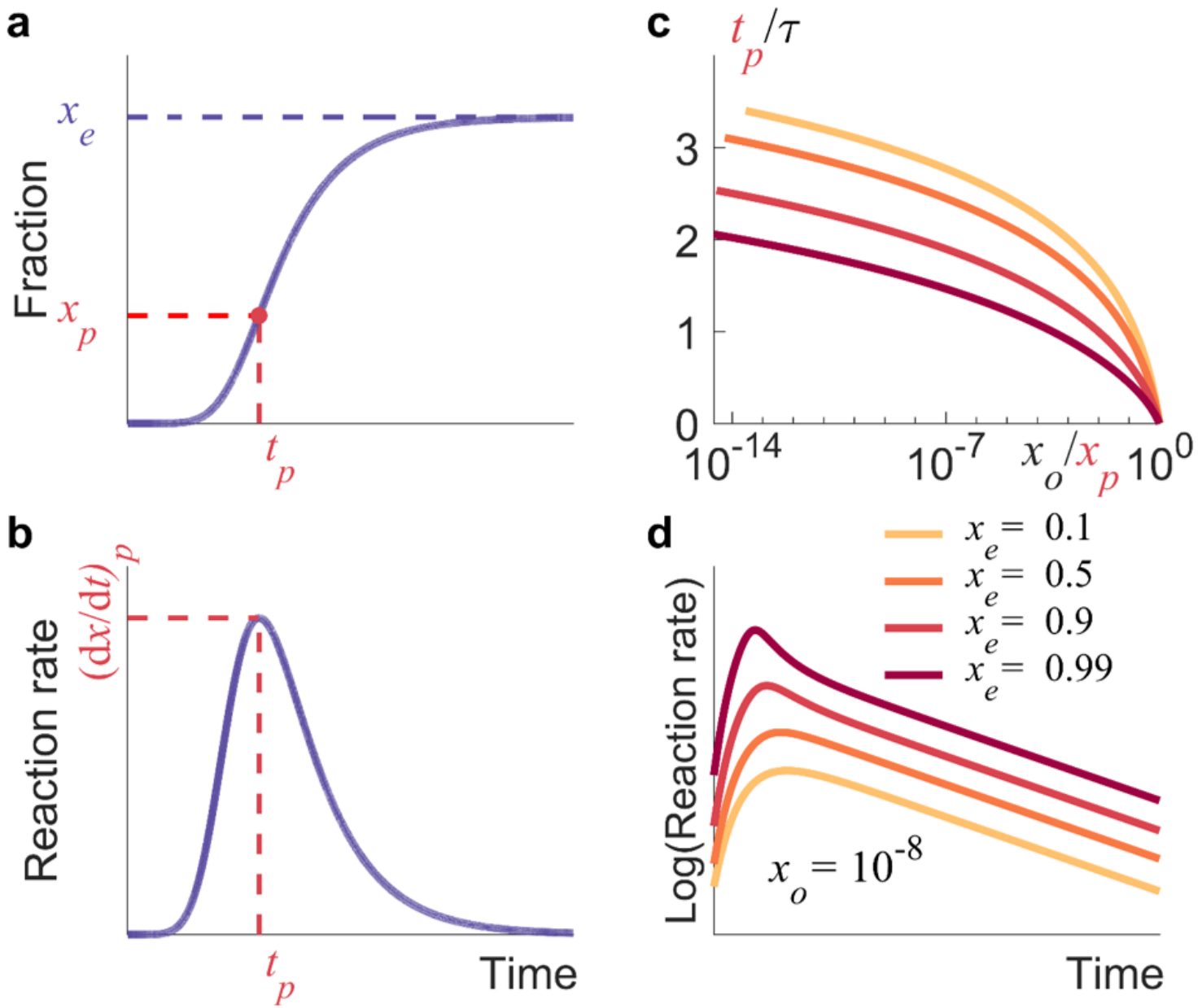}}
\caption{\textbf{Affinity relaxation theory (ART) of a binary reactive mixture} (a), Reaction product fraction when the initial fraction is smaller than $x_p$. (b), Reaction rate showing a peak at time $t_p$. (c) When the initial fraction $x_o$ is smaller than $x_p$, the time at which the peak appears depends crucially on the initial fraction $x_o$. (d) The corresponding overall shape of the rate for a given initial fraction $x_o$ depends on the equilibrium fraction $x_e$ (curves are shifted vertically for better visualization and shown in vertical log scale).}
\label{Fig_ART}
\end{figure}

We aim now to build a thermodynamic model that predicts a reaction rate compatible with the relaxation dynamics observed in epidemic data (Fig.~\ref{FigDataSIR}c). For simple chemical reactions, first order mass-action rate laws would lead to an exponential approach of the reaction rate towards the final state \cite{Laidler1987}. However, at time scales smaller than the relaxation time, the chemical potential convergence would be hyperbolic \cite{SupMat} and not exponential. Alternatively, we could use an Onsager approach with a reaction rate proportional to the thermodynamic force \cite{Groot1984}. For the reactive binary mixture, this thermodynamic force, or affinity \cite{Prigogine1948}, is equal to the difference in chemical potential between the two mixed molecular species. However, this approach also leads to an exponential relaxation of the chemical potential at long times only and again a hyperbolic convergence at shorter times \cite{SupMat}. Instead, we will assume that affinity is a decreasing exponential function of time, extending far away from equilibrium a prediction demonstrated only close to equilibrium \cite{Prigogine1954}. Noting $x$ the fraction of infected people in an initial population $N_o$ of susceptible people and $x_e$ the fraction in the final state where the thermodynamic force becomes zero (equilibrium fraction in a chemical reaction), the affinity $~A$ of the reaction $S \rightarrow C$ is \cite{SupMat}:
\begin{equation}
\beta ~A=\ln \left(\frac{x_e}{1-x_e}\frac{1-x}{x}\right)
\label{affinity}
\end{equation}
where $\beta=1/k_B T$ (Boltzmann constant $k_B$ and temperature $T$). An exponential relaxation of affinity towards equilibrium $~A=~A_o e^{-t/\tau}$ would then lead to the following time evolution of the infected fraction $x$:
\begin{equation}
\frac{x}{1-x}=\left(\frac{x_o}{1-x_o}\right)^{e^{-t/\tau}}\left(\frac{x_e}{1-x_e}\right)^{1-e^{-t/\tau}}
\label{Gompertz}
\end{equation}
where $x_o$ is the initial fraction. This prediction has a mathematical form that falls into the family of Gompertz functions \cite{Winsor1932,Castorina2006,Lanteri2020,Ohnishi2020} used empirically for many growth problems in biology, including epidemic growth. Generally, the time evolution predicted by Eq.~(\ref{Gompertz}) (Fig.~\ref{Fig_ART}a) displays a peak rate $(dx/dt)_p$ at some time $t_p$ (Fig.~\ref{Fig_ART}b). However, the existence of this peak and the time at which it appears to depend both on the initial fraction $x_o$ and the final state fraction $x_e$ \cite{SupMat}. First, the fraction $x_p$ at which the peak rate occurs depends only on $x_e$. Hence, for an initial fraction $x_o$ larger than $x_p$, the peak rate would occur at the start of the reaction. In that case, theory predicts almost a perfect exponential evolution of the reaction rate, as in classical mass-action rate laws. For an epidemic, if we consider that the initial state is only one infected person out of a very large population, the initial fraction $x_0$ will be very small ($\sim 10^{-8}-10^{-6})$, and the peak delay $t_p$ will be of the same magnitude as the relaxation time (Fig.~\ref{Fig_ART}c).
%Th result suggests that the very small initial fraction of infected people is the main reason why there is an observable time for an epidemic to grow and reach a peak rate, before slowing down.

%In the case of a chemical reaction, following the reaction dynamics with molecular fractions varying by 5 to 6 orders of magnitude as can be done for an epidemic would be quite challenging.

\begin{table}
\begin{ruledtabular}
\begin{tabular}{cccc}
& $x_e$ & $r_e$ & $R_0$\\
\hline
Belgium & $0.84\pm0.02$ & $5.25 \pm 0.61$ & $5.00\pm0.73$\cite{Linka2020}\\
%\hline
Canada & $0.83\pm0.02$ &$4.88 \pm 0.74$ & $4.15\pm1.15$ \cite{Krkosek2021}\\
%\hline
Germany & $0.87\pm 0.01$ & $6.69 \pm 0.76$&$6.33\pm0.64$\cite{Linka2020}\\
%\hline
Italy &$0.74\pm 0.06$ & $2.84\pm 0.87$ & $3.27\pm0.11$\cite{Billah2020}\\
%\hline
Netherlands & $0.84\pm0.01$ & $ 5.25\pm0.50 $&$5.88\pm0.88$\cite{Linka2020} \\
%\hline
UK & $0.67\pm 0.02$ & $2.03\pm 0.19$ & $2.1 \pm 0.2$\cite{Lonergan2020}
\end{tabular}
\end{ruledtabular}
\caption{\textbf{Final fraction and probability ratio of the infected population.} Final fraction $x_e$ and probability ratio of being infected $r_e$ in the thermodynamic model compared to the basic reproduction number $R_0$ during the first COVID-19 wave of a few countries.}
\label{TableHIT}
\end{table}

\begin{figure}[htbp]
\centerline{\includegraphics[width=8.5cm]{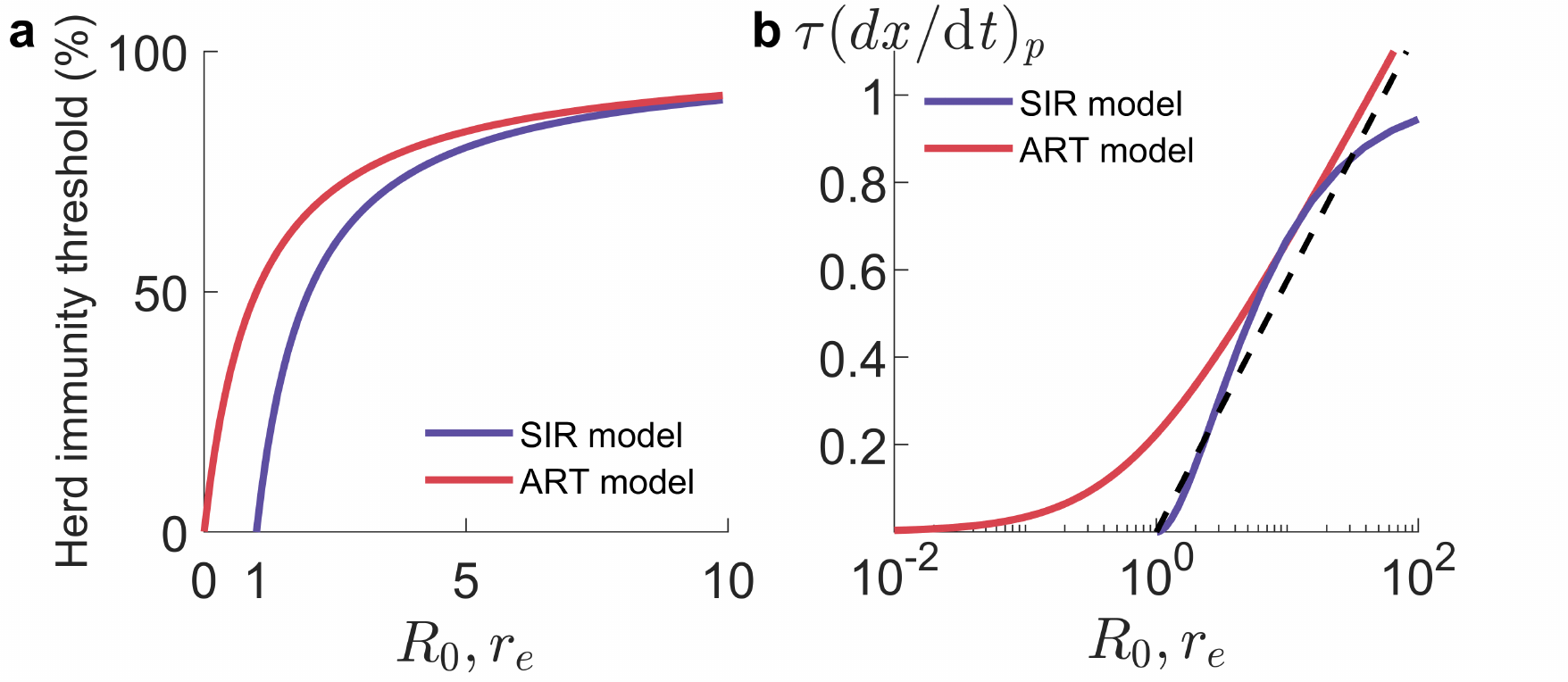}}
\caption{\textbf{Herd immunity and peak rate transition towards epidemic outbreaks} (a), The herd immunity threshold in the ART model is always above the SIR model prediction, and exists for values $r_e<1$. (b), Above $r_e=1$, the peak rate in the ART model increases significantly faster with $r_e$ than below. The peak rate in the SIR model evolves around the asymptotic limit of the ART model for highly contagious epidemic  (dashed line). }
\label{Fig_HIT}
\end{figure}

Starting from an initially very small fraction of infected people is not sufficient to produce a reaction rate shape that agrees with epidemic data. Indeed, the shape depends on the value of the fraction $x_e$ reached in the final state (Fig.~\ref{Fig_ART}d). We estimated the optimum value for $x_e$ by considering the first COVID-19 wave in several countries \cite{owidcoronavirus}, setting the initial fraction at one infected case $x_0=1/N_0$ (Table~\ref{TableHIT}). In a binary reactive mixture, the equilibrium fraction:  $x_e=r_e/(1+r_e)$ is determined by the ratio $r_e=\exp(-\beta\mu^0_C)/\exp(-\beta\mu^0_S)$ between the Boltzmann factors associated with the chemical potential of each molecule in a pure solution, $\mu^0_S$ and $\mu^0_C$ \cite{SupMat}. The ratio $r_e$, otherwise known as the equilibrium constant of the chemical reaction, measures how much more likely it is to find a molecule in state C rather than in state S. This elementary probability ratio carries statistical information somewhat equivalent to the basic reproduction number $R_0$ in epidemic models. This equivalence is reinforced by comparing the values of $r_e$ associated with $x_e$ to $R_0$ estimates from the literature \cite{Linka2020,Krkosek2021,Billah2020,Lonergan2020} (Table~\ref{TableHIT}). In the thermodynamic model, the fraction $x_e$ towards which an epidemic wave would converge and stop may be interpreted as the fraction of infected people required to reach herd immunity. However, in the SIR model, the herd immunity threshold $x_H$ has a different meaning. It is obtained from the basic reproduction number $R_0$ by stating that the effective reproduction number $R_e\equiv R_0 (1-x) = 1$, which occurs at the peak in infectious population I, just before the epidemic wave starts to slow down and well before it ends. Nevertheless, this has been used as a criterion for herd immunity leading to a herd immunity threshold $x_H = 1 -1 /R_0$ \cite{Diekmann2012}. In comparison, the thermodynamic model predicts, for large values of $r_e$, a herd immunity threshold $x_e \simeq 1 - 1/r_e$. Thus, in the limit of highly contagious viruses, the thermodynamic criterion for herd immunity is mathematically the same as in the SIR model. However, for smaller values of $r_e$, the two models differ (Fig.~\ref{Fig_HIT}a). A major difference is that the herd immunity criterion in the SIR model applies only to values $R_0>1$, while the criterion in the thermodynamic model always applies, even if $r_e<1$. Interestingly, some viruses have been estimated to have a basic reproduction number $R_0<1$. For instance, MERS ($R_0 \simeq 0.47$) \cite{Kucharski2015} and Nipah virus ($R_0 \simeq 0.48 $) \cite{Luby2013} have produced small epidemic outbreaks, for which the thermodynamic model predicts a herd immunity $x_e \simeq 32\%$ (taking $r_e \approx R_0$). Generally, the herd immunity criterion in the SIR model underestimates the herd immunity criterion of the thermodynamic model (e.g., for the Andes virus $R_0\simeq 1.2$ \cite{Martinez2020}, $x_e=54\%$ against $x_H=16\%$). Another important result is that the peak rate predicted by the thermodynamic model increases with $r_e$ (Fig.~\ref{Fig_HIT}b). While the increase is slow and peak rate values remain small for $r_e<1$, the increase becomes more noticeable for $r_e>1$, with peak rate values asymptotically diverging as $\ln r_e/4 $ (dotted line in Fig.~\ref{Fig_HIT}b). As a consequence of this peak rate transition, epidemic waves that would occur above $r_e=1$  will tend to be large since their peak rate is high, while below $r_e=1$, epidemic waves that would not occur in the SIR model, will occur in the thermodynamic model but remain small since their peak rate is low. The SIR peak rate, with time rescaled by the relaxation time $\tau_R=\gamma^{-1}$, approximately follows the same evolution as the asymptotic regime of the thermodynamic model (Fig.~\ref{Fig_HIT}b). Notably, the asymptotic peak rate is reached when 50$\%$ of a population has been infected, which is also when mixing entropy is maximum. As a consequence, during the slowing-down phase of the epidemic wave, a significant part of the population still needs to be infected to reach thermodynamic herd immunity.

For the six countries considered, the relaxation time has an average value $\tau=27.7 \pm 5.2$ days \cite{SupMat}. The similar exponential decrease observed in the SIR model at long times (Fig.~\ref{FigDataSIR}d) occurs at a rate that is exactly the recovery rate of the infectious population \cite{SupMat}. However, the rather large values obtained for $\tau$ suggest that the relaxation time cannot be simply equated with the recovery time. While recovery time is an elementary time for the epidemic process at the scale of an individual, the relaxation time in the thermodynamic model is the result of a collective relaxation process at large scale involving mixing entropy and may depend on  parameters other than the recovery time such as the speed of virus circulation and population density, size and heterogeneity.

\begin{acknowledgments}
L. Vanel thanks S. Ciliberto for scientific discussions.
\end{acknowledgments}

\bibliographystyle{apsrev4-2}
\bibliography{citations.bib}

\end{document}